\documentclass{mem}
\usepackage{natbib}\usepackage{txfonts}\usepackage{balance}
\usepackage{graphicx}
\usepackage[a4paper]{hyperref}

\idline{75}{282}
\begin{document}
\def\teff{$T\rm_{eff }$}
\def\kms{$\mathrm {km s}^{-1}$}

\title{
The exciting future of (sub-)millimeter interferometry: ALMA
}

\subtitle{}

\author{
V. \,Casasola 
\and J. \,Brand
}

\offprints{V. Casasola}

\institute{
INAF -- Istituto di Radioastronomia \& Italian ALMA Regional Centre \\ 
Via P. Gobetti 101, 40129 Bologna, Italy \\
\email{casasola@ira.inaf.it; brand@ira.inaf.it}
}

\authorrunning{Casasola \& Brand}

\titlerunning{ALMA}

\abstract{
The Atacama Large Millimeter/submillimeter Array (ALMA), presently under construction, 
is a revolutionary astronomical interferometer, that will operate 
at (sub)millimeter wavelengths.
With unprecedented sensitivity, resolution, and imaging capability, 
ALMA will explore the (sub-)mm Universe, one of astronomy's last frontiers.
ALMA is expected to provide insight in star- and galaxy formation in the early 
Universe and to image local star- and planet formation in great detail.
The ALMA Commissioning and Science Verification phase is currently in course, 
preparing the path for Early Science. The Call for ALMA Early Science 
proposals is expected to be released before the end of 2010. 
In this contribution we will describe the ALMA project, the array and its 
receivers, its science goals, and its scientific and technological potential.
We will outline the organizational structure of the ALMA Regional Centres, 
that will play an important role in providing support to the users, with 
particular attention to the Italian ALMA Regional Centre in Bologna. 
Finally, we will illustrate what ALMA can contribute to the specific 
science case of AGN fueling.
\keywords{Instrumentation: interferometers -- Instrumentation: high angular 
resolution -- Galaxies: individual: NGC\,5953 -- Galaxies: active -- 
Galaxies: nuclei}
}
\maketitle{}

\section{Introduction}
In essence, the Atacama Large Millimeter/submillimeter Array (ALMA)
is a radio interferometer, that because of its design {\it and} its unique location will 
allow us to perform ground-breaking science in the field 
of millimeter (mm) and submillimeter (sub-mm) astronomy.
ALMA, thanks to its unprecedented sensitivity and resolution, will open up a new 
window onto the cold Universe, 
capturing never-before seen details about the very first stars and 
galaxies in the Universe, and directly imaging the formation of planets. 

In Sect. \ref{sec:coll-site}, we briefly describe
the  global collaboration behind ALMA and the present status of the project. 
The main technical specifications and the primary science goals of ALMA are presented 
in Sects. ~\ref{sec:tec-spe} and \ref{sec:science}, respectively. Early Science observations 
are described in Sect.~\ref{sec:obs}. 
In Sect.~\ref{sec:arc} we outline the organizational stucture of the ALMA Regional 
Centres, and their role in providing support to future users with proposal preparation and 
post-observation data reduction. Particular attention will be given to the Italian node of the 
network. Finally, Sect. \ref{sec:nuga}
shows a specific example of a science project, and how this will benefit from 
observations with ALMA.

\section{ALMA: A global collaboration \label{sec:coll-site}}
ALMA is designed to explore the cold Universe by means of observations of 
mm and sub-mm radiation, at wavelengths between 0.3 and 3~mm.
Since in the (sub-)mm range the signal from space is heavily absorbed 
by water vapour in the Earth's atmosphere, instruments operating at these wavelengths must 
be built on high and dry sites. For ALMA, the 5000\,m high plateau at Chajnantor has been  
chosen, one of the highest astronomical observatory sites on Earth.
The ALMA site, $\sim$50\,km east of San Pedro de Atacama in 
northern Chile, is also one of the driest places on Earth.

ALMA is a truly global endeavour, a partnership between Europe, North America 
(USA and Canada), and East Asia (Japan and Taiwan), in cooperation with the 
Republic of Chile. 
The project development is coordinated by the Joint ALMA Office (JAO), based in 
Santiago de Chile.

The ALMA Observatory will operate at two distinct sites: 
the Operations Support Facility (OSF) and the Array 
Operations Site (AOS). The AOS lies  at an altitude of 5000~m and is where 
the array and correlator will be located when ALMA is in operation. 
The OSF is situated at $\sim$2900\,m altitude and will be the base camp 
for the routine operations of the observatory. 
The OSF is the focal point of all antenna 
Assembly-Integration-Verification (AIV) activities. During the AIV-stage the antennas are
assembled, integrated with the other subsystems (electronics, receivers), and tests are 
performed to ensure that the system satisfies the astronomical 
requirements. The antennas are then transported to the AOS, 28~km 
away and 2100~m higher. With the 
delivery of 3 fully equipped and functioning antennas to the high site in January 2010, 
the Commissioning and Science Verification (CSV) stage began. In short, the goal of the 
CSV is to test the operation of the ever-growing array and to obtain quantitative confirmation 
that the data 
have the required characteristics in terms of sensitivity and image quality. 


The day-by-day operation of the observatory and all maintenance will take place 
at the OSF. 
During the operations phase of the observatory, the OSF will be the workplace of 
the astronomers and of the teams responsible for maintaining the proper 
functioning of all the telescopes. 
The quality of all ALMA data will be assessed at the OSF. 
In other words, the OSF will be, in many aspects, the centre of activities of the 
ALMA project, while human operations at the AOS will be limited to an 
absolute minimum, due to the high altitude.


The development of ALMA is proceeding rapidly: preparation of the site started in late 
2005, and construction will be completed in 2013. The first prototype antenna was tested 
(at the VLA in Socorro) in 2003; first fringes between two antennas were 
detected there in March 2007, and a month later the first ALMA production antenna arrived in 
Chile. In mid-2009 two-antenna interferometry was done at the OSF, and by the end of 
that year 3-antenna interferometry and phase closure was achieved at the AOS. Commissioning 
and Science Verification started in early 2010. Presently there are 8 antennas at the 
high site.

\begin{table*}
\caption{Main parameters of ALMA receiver bands.}
\label{bands}
\begin{center}
\begin{tabular}{llllllll}
\hline
\\
Band & Frequency & Wavelength & Angular & Line & 
Continuum & Primary & Largest \\
& range & range & resolution & sensitivity & 
sensitivity & beam & scale \\
& &  & b$_{\rm max}=$\,200\,m--16\,km &  &  &  &  \\
& [GHz] & [mm] & [$^{\prime\prime}$] & [mJy] & [mJy] & [$^{\prime\prime}$] & [$^{\prime\prime}$] \\
\hline
\\
3  & 84--116  & 2.6--3.6 & 3.0--0.034 & 8.9 & 0.060 & 56 & 37 \\
4  & 125--169 & 1.8--2.4 & 2.1--0.023 & 9.1 & 0.070 & 48 & 32 \\
5  & 163--211 & 1.4--1.8 & 1.6--0.018 & 150 & 1.3   & 35 & 23 \\
6  & 211--275 & 1.1--1.4 & 1.3--0.014 & 13 & 0.14   & 27 & 18 \\
7  & 275--373 & 0.8--1.1 & 1.0--0.011 & 21 & 0.25   & 18 & 12 \\
8  & 385--500 & 0.6--0.8 & 0.7--0.008 & 63 & 0.86   & 12 & 9  \\
9  & 602--720 & 0.4--0.5 & 0.5--0.005 & 80 & 1.3    & 9  & 6  \\
\hline
\end{tabular}
\end{center}
\end{table*}

\section{ALMA technical specifications \label{sec:tec-spe}}
The ALMA interferometer operates at wavelengths between 
0.3 and 3.0\,mm (84 to 720~GHz). 
It will be composed initially of 66 antennas, 
with a possible extension in the future. 
The main array is constituted of  fifty 12-m diameter antennas, 
In addition, twelve 7-m diameter antennas for interferometry and 
four 12-m diameter antennas for total power observations form the Atacama 
Compact Array (ACA). 
The ACA is used to obtain short spacing data.
Because two antennas can not be placed closer than some minimum 
distance ($D_{\rm min}$ = 15~m for the ALMA array), signals on spatial scales larger 
than some size ($\propto$$\lambda/D_{\rm min}$) will not be detected. 
This effect, called the ``missing flux'' problem, is resolved by observations with the
ACA and by then combining the ACA-data with the main array measurements.
Therefore, the ACA works together with the main array in order to enhance 
the wide field imaging capability.

The main array can be configured in various ways, ranging in size between 200\,m and 
16\,km, in order to achieve specific imaging requirements.
Compact configurations are indicated to image extended sources because
they are more sensitive to low surface brightness features, 
while extended configurations 
allow one to achieve a better resolution (at the expense of surface brightness sensitivity).
ACA on the other hand, will have essentially one configuration ($\sim$50~m in size) 
with the possibility of 
a slight north-south extension to reduce blockage of the inner antennas by the other 
ones when observing sources at northern declinations.

Table~\ref{bands} collects the main parameters of ALMA receiver bands.
In this table, Col. (1) indicates the number of the ALMA receiver band; 
Cols. (2) and (3) the frequency and wavelength ranges, respectively, 
covered by each band; Col. (4) the angular resolution for the most compact and the 
most extended configurations; 
Cols. (5) and (6) the line and continuum sensitivities, respectively; 
Col. (7) the primary beam size; and Col. (8) the largest observable scale 
in a given band.

The frequency range available to ALMA is divided into 10 receiver bands, 
that can be used only one band at a time.
ALMA will operate initially in 7 bands: from band 3 to 9 (see Table~\ref{bands} for a list 
of the main parameters). 
Bands 1, 2, and 10 at 
$\sim$40\,GHz (7.5\,mm), $\sim$80\,GHz (3.7\,mm), and $\sim$920\,GHz
(0.3\,mm), respectively, might be added in the future.

The spatial resolution depends on the observing frequency and 
the maximum baseline of the array. 
In the most compact ALMA configurations (200\,m), the spatial resolution 
ranges from 0\farcs4 at 675\,GHz to 2\farcs8 at 110\,GHz, while in the most 
extended configuration, it ranges from 6\,mas at 675\,GHz to 38\,mas 
at 110\,GHz. 
The receivers have an instantaneous bandwith of 8~GHz, that can be divided in 
up to 4 spectral windows of up to 2~GHz each.
ALMA can produce data cubes with up to 8192 frequency channels, whose width may 
range between 3.8\,kHz and 2\,GHz. 

The field of view (FOV) is determined by the size of a single antenna and 
by the observing frequency, but is independent of the array configuration. 
The FOV is expressed in terms of the primary beam, which describes 
the antenna response (sensitivity) as a function of the angle away from the 
main axis. 
At 300\,GHz, the FWHM of the ALMA primary beam is 17$^{\prime\prime}$; 
this parameter scales linearly with wavelength.

In interferometry, the signals of all individual antennas that make up an array are 
combined, or correlated, in order to simulate an observation with a single-dish 
telescope that has the size of the array. It is important that all signals have the same 
phase, otherwise the correlated signal will have no output. Very careful timing of, and 
correction for, the delay with which signals from different antennas are combined is thus of the essence. Water vapour in the atmosphere slows down the propagation of the signals, 
making them arrive slightly later than they would have without water in the atmosphere. 
This is especially problematic if the distribution of water vapour is non-uniform (and thus 
varies from antenna to antenna). Furthermore, the amount of water vapour varies with time.
Observations in the (sub-)mm regime are particularly susceptible to this, especially at 
the high-frequency range.  
Therefore,  in order to successfully correlate the signals from the various antennas, one 
has to know the amount of water vapour in the atmosphere. This is achieved by using a 
``water vapour radiometer'', which is installed at every antenna, and monitors the amount 
of water vapour in the atmosphere by measurements at 183~GHz. These data are then 
used to correct for the delay in the arrival times of the signals.

\section{ALMA science goals \label{sec:science}}
The design of ALMA and the technical specifications outlined in the previous 
section, are based on three main key science goals:
\begin{enumerate}
\item
the ability to detect spectral line emission from CO and CII in normal galaxies at 
redshift $z = 3$ in less than 24 hours of observation;

\item
the ability to image the gas kinematics in protostars and in protoplanetary disks 
around young Sun-like stars at a distance of up to 500~light years, to study their 
physical, chemical, 
and magnetic field structures, and detection of the tidal 
gaps created by planets undergoing formation in the disks;

\item
to provide images at an angular resolution of better than 0.1$^{\prime\prime}$,
and at high dynamic range.
\end{enumerate}

In addition to these three main key science goals, 
other important topics will be explored with ALMA, such as:
\begin{itemize}

\item
redshifted dust continuum emission from evolving galaxies 
at epochs of their formation as early as z = 10.
The inverse K-correction will make ALMA the ideal instrument for 
investigating the origin of the galaxies in the early Universe, 
with confusion minimized by the high spatial resolution;

\item
CO emission to derive the redshift of star-forming galaxies;

\item
cold dust and molecular gas in nearby galaxies to study the 
interstellar medium in different galactic environments, 
the effect of the physical conditions on the local star formation 
history, and galactic structure;

\item
kinematics of obscured AGN and quasars on spatial scales of 
10-100\,pc, to test unification models of Seyfert galaxies;

\item
dynamics of the molecular gas at the centre of our own Galaxy 
with unprecedented spatial resolution revealing the tidal, 
magnetic, and turbulent processes;

\item
detailed analysis of how stars form from the gravitational 
collapse of dense cores in molecular clouds, and of the 
formation and evolution of disks and jets in young protostellar 
systems;

\item
formation of molecules and dust grains in the circumstellar 
shells and envelopes of evolved stars, novae, and supernovae;

\item
refinement of dynamical and chemical models of the atmospheres 
of planets in our own Solar System, and imaging of cometary nuclei, 
hundreds of asteroids, Centaurs, and Kuiper Belt Objects. 

\end{itemize}

The Italian astronomical community has strong interests in several of these lines 
of research.


\section{Early Science observations \label{sec:obs}}
At the moment of writing (October 2010) there are 8 antennas at the AOS, 
and new antennas and receivers will follow at regular intervals.  
It will not be necessary to wait for all 66 (50+16) antennas to 
be at the AOS to do science. When certain minimum requirements are 
met, the project enters into its next phase, that of Early Science. 

Early Science can be done when the following conditions apply:
\begin{itemize}

\item
at least 16 12-m antennas fully commissioned;

\item
array configurations sufficient to cover the shortest 
spacings and out to a maximum baseline of 250\,m (but 
hopefully 1\,km); 

\item
at least three frequency bands on each antenna 
(hopefully four: bands 3, 6, 7, and 9; possibly plus 
bands 4 and 8 on as many antennas as possible);

\item
single-field interferometry observing mode 
(and hopefully interferometric pointed mosaics);

\item
a mixture of continuum and spectral line correlator configurations;

\item
calibration to a level already achieved on established mm arrays;

\item
tools required for proposal submission, preparation and execution of 
observations, and data reduction are in place.

\end{itemize} 

The decision to issue a Call for Proposals for Early Science will be made when 
it is likely that the above-listed requirements will be met within the next eight months.
This Early Science Decision Point (ESDP) is expected to be early December 2010. 
The deadline for submitting proposals will be two months after the ESDP and observations 
will begin six months after that.

In general, when the Joint ALMA Office (JAO) issues calls for proposals, 
an astronomer who wishes to apply for observing time, will have to 
register on the ALMA web page. 
The proposal will be prepared with the ALMA Observing Tool (AOT). 
The AOT, a java application, is a software package 
to construct a proposal, called Observing Project. 
This Observing Project consists of two parts: 
a Phase I Observing Proposal containing the scientific justification 
of the proposed observations and a minimal amount of technical 
information required to check the feasibility of the proposal, 
and a Phase II Observing Programme, to be submitted only if observing 
time has been granted. 
The JAO, with assistance from the ALMA Regional Centres (ARCs)
[see Sect.~\ref{sec:arc}], coordinates the refereeing process of the 
submitted proposals. There will be a single Time Allocation Committee.

In the case of a successful proposal, the user will be required to 
prepare the Phase II programme, providing full details of the proposed 
observations. The user will not 
travel to Chajnantor to carry out the observations, but the programmes  
will be entered into a queue to be dynamically scheduled, depending on 
requested weather conditions and array configuration.
The data obtained will pass through a quality assurance programme
(such as on-site checks by the astronomer-on-duty, a quick-look analysis, system 
performance checks), will then go through the data reduction 
pipeline, and be delivered to the archive.
P.I.'s of the proposals will be notified immediately after their data become 
available. They will receive the pipeline products, such as fully calibrated 
images or data cubes, and calibrated u-v data. 

The ALMA data will be reduced with the Common Astronomy Software 
Applications package, better known as CASA.
CASA has been developed with the primary goal to support the 
data post-processing needs of the next generation of radio  
telescopes, such as ALMA but also the Expanded Very Large Array (EVLA).
CASA can process both interferometric and single-dish data, 
and is developed by an international consortium 
led by the 
National Radio Astronomical Observatory (NRAO). 
CASA consists of a set of C++ tools bundled together under 
an iPython interface as a set of data reduction tasks. 
This structure provides flexibility to process the data via 
a task interface or as a python script. 
In addition to the data reduction tasks, many post-processing tools 
are available for even more flexibility and special purpose 
reduction needs. 
CASA is currently in Beta (3.1) release, and a Cookbook and Reference Manual 
are available from the NRAO CASA web site (http://casa.nrao.edu/).

\section{The Alma Regional Centre network \label{sec:arc}}
ALMA is intended to be an instrument for which even a non-expert in mm-interferometry 
should be able to write a successful proposal. Much attention is therefore given 
to user support. 
The interface between ALMA and the user community is managed 
by three ALMA nodes, called ALMA Regional Centres (ARCs),
located in Europe, North America, and East Asia.
The European ARC is at ESO, Garching, 
and serves as the access portal to ALMA for the European user community.
The EU ARC manages a network of local European ARC nodes,  
which provide additional services to the whole of the EU community. The details of the 
responsibilities of the various partners in the network are described in a Memorandum of 
Understanding.
The European ARC-nodes are located at Bonn-Bochum-Cologne (Germany), 
Bologna (Italy), Onsala (Denmark, Sweden, and Finland), IRAM-Grenoble (France), 
Leiden (The Netherlands), Manchester (United Kingdom), and Ondrejov (Czech Republic).
The support offered by the ARCs to the user community includes:
\begin{itemize}
\item
archiving and data reduction support, such as face-to-face data 
processing support, modified pipeline versions, re-processing 
of large and/or complex datasets, simulation development, and 
help with archival research projects;
\item
support for special projects, such as public surveys;
\item
science community development, such as support for ALMA research, 
post-doctoral fellowships, and specialized schools and workshops.
\end{itemize}

Each local ARC node contributes with its own specific expertise, 
in order to ensure that maximum advantage of the European 
competences in the field of (sub-)mm interferometry is taken.
More information on the European ARC network is available on the web pages 
http://www.eso.org/sci/facilities/alma/arc/, where links to its  
North American and Japanese counterparts can also be found.

\subsection{The Italian ARC}
The Italian ARC-node is hosted by the Istituto di Radioastronomia in Bologna, and is 
funded by INAF.
When ALMA is operational, the ARC-node will have 1 staff-member and 4 post-docs 
that spend about 50$\%$ of their time to provide user support, including face-to-face 
help, and a system manager who takes care of the software and hardware 
of the ARC (which has a dedicated server), and in particular with the development of the  
of GRID technology. Furthermore, from October 2010 an ESO-ALMA Fellow will join the 
ARC for 3 years, who will be especially involved with the scientific part of the preparation 
of programmes for Early Science. 
The members of the Italian ARC are experienced in the reduction of interferometric data 
with CASA and in the use of the AOT. They regularly participate in tests of the various 
software packages, organized on an international level. 
To inform the astronomical community we regularly organize 
so-called ALMA-days, in which updates are given of the ALMA project and experts are 
invited to talk about the impact of ALMA on various fields of research. To instruct the 
community in the use of the CASA data reduction package we have recently organized 
a CASA-tutorial. In the Summer of 
2011 the ARC will organize, in collaboration with the EU COST-Action network, an 
international school on ``Astrochemistry with ALMA''.

When ALMA users need assistence in preparing observing proposals, when they 
require support with the reduction of their data 
(if the pipeline products are not to their satisfaction), then they can contact the central 
helpdesk at ESO, which will then either answer the questions or direct the user to one of 
the ARC nodes if specialist help is required. The various nodes in the EU-ARC network 
have particular areas of expertise, and provide face-to-face help to users who require this.
The Italian ARC offers specialized help with polarimetry, mosaicking, and the handling of  
large data sets.
Around the time of the Call for Proposals for Early Science a ``Guide to the 
European ARC'' will be published, which details the type of support offered, and how to get 
access to it. 
For more information please visit our website http://www.alma.inaf.it.

\begin{figure*}
\centering
\includegraphics[width=0.43\textwidth,angle=-90,bb=145 17 460 750]{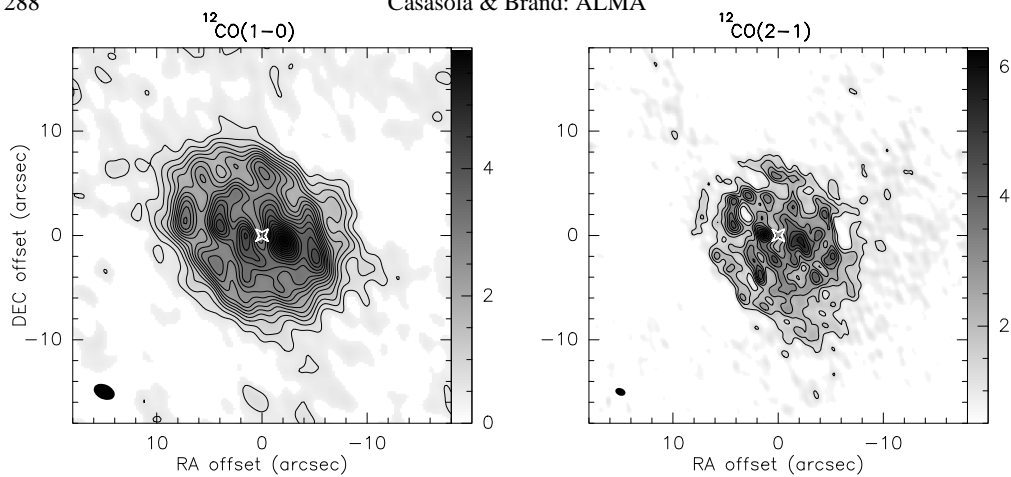}
\caption{
\footnotesize
\textit{Left panel}: $^{12}$CO(1--0) integrated intensity contours 
observed with the IRAM PdBI+30-m towards the centre of NGC\,5953. 
The white star marks the coordinates 
of the dynamical centre of the galaxy, with offsets 
in arcseconds.
The map, derived with 2$\sigma$ clipping, has not been corrected for
primary beam attenuation.
The {\it rms} noise level is $\sigma = 0.09\,{\rm Jy\,beam^{-1}\,km\,s^{-1}}$ 
and contour levels run from 3$\sigma$ to 21$\sigma$ with 3$\sigma$ spacing.
In this map a $\pm 130\,{\rm km\,s^{-1}}$ velocity range is used. 
The beam of 2\farcs1 $\times$ 1\farcs4 is plotted at lower left.
\textit{Right panel}: Same for $^{12}$CO(2--1).
The {\it rms} noise level is $\sigma = 0.2\,{\rm Jy\,beam^{-1}\,km\,s^{-1}}$ and 
contour levels run from 3$\sigma$ to 10$\sigma$ with 3$\sigma$ spacing.
The beam of 1\farcs1 $\times$ 0\farcs7
is plotted at lower left.
Figure from \citet{vivi10}.
}
\label{n5953}
\end{figure*}

\section{An example of science in preparation of ALMA \label{sec:nuga}}
In this section, we discuss a specific ALMA science case, 
the AGN fueling, its state of the art, and what we will learn about 
this topic with ALMA.
Identifying the main mechanism responsible for fueling 
AGN has been one of the hottest topics in the last decade. 
Non-axisymmetric potentials introduced by stellar bars 
and galaxy interactions were considered as promising mechanisms
for fueling local low-luminosity AGN (LLAGN).
CO observations offer the possibility of directly witnessing the gas
fueling towards the nucleus.
Observational campaigns have been performed with millimeter instruments 
to investigate this issue: the NUclei of GAlaxies (NUGA) survey, performed with 
the IRAM Plateau de Bure Interferometer (PdBI) and dedicated to a sample of 12 nearby LLAGN, 
found that these objects are characterized by a wide variety of 
molecular gas distributions (e.g., streaming motions along stellar
bars, rings, nuclear concentrations, nuclear voids, and irregular distributions), 
and that sometimes the gravitational torques exerted by the stellar potential on 
those gas distributions can fuel the central AGN 
\citep[e.g.,][]{santi03,francoise04,vivi08,vivi10}.
This suggests that several mechanisms, rather than 
a single universal one, can cooperate to fuel
the central engines of LLAGN.

Fig. \ref{n5953} shows $^{12}$CO(1--0) and $^{12}$CO(2--1) 
integrated intensity distributions obtained for the interacting Seyfert 2/LINER galaxy 
NGC\,5953 by combining IRAM PdBI and single-dish 30-m observations, in the context of 
the NUGA project.
The CO emission is distributed over a disk of diameter of $\sim$16$^{\prime\prime}$.
Our $^{12}$CO(1--0) observations show several peaks, distributed more or less 
randomly, with the strongest one offset from the nucleus by $\sim$2$^{\prime\prime}$.
In the $^{12}$CO(2--1) map the central emission is also clearly resolved and 
more clumpy than in $^{12}$CO(1--0).
The strongest $^{12}$CO(2--1) peak is not that at $\sim$2-3$^{\prime\prime}$ 
to the W/SW from the nucleus, like for $^{12}$CO(1--0), 
but that at $\sim$1\farcs5 E from the nucleus.

High-resolution optical and near-infrared images of galaxies
allow one to quantify the efficiency of the stellar potential 
in draining angular momentum in a galaxy by deriving torques 
exerted by the potential on the gas.
In NGC\,5953, we found that the torques are predominantly
positive in both $^{12}$CO(1--0) and $^{12}$CO(2--1), 
suggesting that gas is not flowing into the centre.
This comes from the regular and almost axisymmetric
total mass and gas distributions in the centre of the galaxy.
In NGC\,5953, the AGN is apparently not being actively fueled 
in the current epoch.

The results obtained from the analysis of stellar torques in NUGA 
galaxies have revealed a puzzling feeding budget in the circumnuclear 
disks. 
Paradoxically, feeding due to the stellar potential seems to be 
presently inhibited close to the AGN for the $\sim$50$\%$ of the analyzed 
galaxies.
\citet{santi05} suggest that gravitational torques could be assisted 
by other mechanisms, such as the viscosity, that become 
competitive with non-axisymmetric perturbations.
Gravitational torques and viscosity could combine to 
produce recurrent episodes of activity during the typical 
lifetime of any galaxy.

Concerning the AGN fueling, ALMA will allow us to improve 
the statistics of the AGN surveys, by increasing the size of 
galaxy samples by orders of magnitude.
The typical ALMA spatial resolution will provide a sharp view
of the distribution and kinematics of molecular gas in the central 
pc of nearby AGN.
Moreover, local AGN have low luminosity and do not require 
high fueling rates from the host galaxies: 
for most local Seyfert nuclei black hole accretions rates are of 
$\sim$10$^{-3}$\,M$_\odot$\,yr$^{-1}$, and at these rates a single 
molecular cloud of 10$^{6}$\,M$_\odot$ can keep the nuclear activity 
going for $\sim$1\,Gyr. 
The fueling problem becomes more serious  for high-luminosity AGN, 
like radio galaxies and quasars, whose black hole accretions rates 
may be $>$1\,M$_\odot$\,yr$^{-1}$.
Quasars are much more distant than Seyfert galaxies, and the limited 
sensitivity and angular resolution of the current millimeter 
interferometers do not allow us to map the molecular gas distribution 
in quasars, except for a few cases. 
With ALMA, instead, we will easily map the gas in quasar hosts.

ALMA will allow us to analyze in detail the bar/AGN feeding cycles of 
distant galaxies and to compare fueling mechanisms as a function of 
redshift.
ALMA will deepen our understanding of how stars form and how star 
formation and AGN fueling processes interact. 
Finally, we will be able to perform Galactic-scale science in distant galaxies.

%

\begin{acknowledgements}
V. Casasola wishes to thank the organizers of the 54$^{\circ}$ SAIt Congress, 
for an interesting and stimulating meeting. We are grateful to Isabella Prandoni for 
a critical reading of an earlier version of the typescript. 
\end{acknowledgements}

\bibliographystyle{aa}

\end{document}